\documentclass[sigconf]{acmart}

\renewcommand\footnotetextcopyrightpermission[1]{}
\acmConference[CHIWork '26]{CHIWork '26 Workshop: Interrogating GenAI Augmentation for CHIworkers}{June 22, 2026}{Linz, Austria}
\acmYear{2026}

\newcommand{\workshopNote}{%
  \begin{quote}
  \small\itshape
  This position paper is submitted to the CHIWork 2026 Workshop:
  \emph{Interrogating GenAI Augmentation for CHIworkers:
  Strategies for Professional Autonomy and Accountability}
  (June 22, 2026, Linz, Austria).
  Workshop proposal: \cite{sandhaus2026interrogating}.
  \end{quote}%
}

\title{Adversarial Co-Thinking: Calibration and Triangulation Across Multiple GenAI Tools in HCI Writing}

\author{Pia Tukkinen}
\affiliation{%
  \institution{Aalto University}
  \city{Espoo}
  \country{Finland}}
\email{pia.tukkinen@aalto.fi}

\renewcommand{\shortauthors}{Tukkinen}

\begin{abstract}
This paper examines what happens when GenAI tools are fully embedded in the drafting of an academic paper rather than confined to late-stage polishing. To investigate how an intensive multi-tool GenAI workflow differs from conventional academic writing, I drafted this paper from the first sentence in parallel with three GenAI tools---Claude, ChatGPT, and Gemini---comparing their outputs against my own intended contribution. Across this process, a recurring pattern took shape that I call \emph{adversarial co-thinking}: using past peer reviews to calibrate the tools, then setting their outputs against one another to be tested rather than deferred to. I argue that surfacing genuine critique from tools that default to praise is a central practical challenge of working with these tools, and that the skill at stake is evaluative rather than generative. Adversarial co-thinking is a high-skill epistemic practice: it can amplify expertise where it exists, but it can also mask its absence. I further argue that current disclosure frameworks are poorly equipped to capture this shift. The paper offers four propositions for workshop discussion concerning autonomy, supervision, equity of access, and disclosure.
\end{abstract}


\begin{document}
\maketitle
\workshopNote

\section{Introduction}

For this paper, I drafted from the first sentence in parallel with three GenAI tools---Claude, ChatGPT, and Gemini---to investigate how an intensive multi-tool workflow differs from writing without one. Unlike my earlier writing, where AI tools entered only at the end---for language refinement and adversarial review of texts I had already drafted---I went all in with AI from the outset for this paper. The paper you are reading is the documentation of that process. This is a first-person methodological account in the tradition of reflexive and autoethnographic HCI research~\cite{desjardins2021firstperson}, in which the researcher's own situated experience is the analytic object and the researcher's interpretation is the analytic instrument. More precisely, this is a retrospective account: the drafting process was treated as data, and this paper is my interpretation of it. The account draws on saved AI drafts, prompt exchanges, my drafting notes, and comparisons between successive versions of this paper. The unit of analysis was the divergence between what I intended to argue, recorded in pre-drafting notes, and what each tool produced or proposed; these divergences---and my responses to them---are what the paper interprets. The polished prose came through the tools; the interpretation, argument, and editorial decisions are mine. The practices described here are a response to a specific configuration of constraints---returning to active research after a long pause, working remotely across time zones, and lacking an in-person research community---and should be read as a situated case rather than a generalizable default.

Out of that process emerged a practice I call \emph{adversarial co-thinking}: calibrating the tools against past peer reviews, then setting their outputs against one another to be tested, rather than deferring to any single tool. The technical components are familiar---few-shot prompting, cross-tool comparison---but the contribution is in their combination and in the evaluative stance the practice requires. This addresses a pressing question for HCI research workflows: how can professionals maintain intellectual autonomy when working with tools that generate plausible but potentially misleading outputs~\cite{sandhaus2026interrogating}? Unlike methodological triangulation, which seeks convergent validation across sources, adversarial co-thinking treats disagreement between tools as the signal---comparison surfaces blind spots that no individual tool would flag on its own. The central skill at stake, then, may not be writing or analysis, but evaluative judgment---distinguishing sound arguments from merely confident-sounding ones.

There is also a more affective dimension to working with these tools. They can partially replace the informal intellectual exchange that researchers in active, in-person groups take for granted---the hallway conversations where vague ideas get tested and refined. For researchers without such a community, this substitution can fill a real gap, but it also introduces risks of its own, most notably a persistent tendency toward sycophantic feedback. Surfacing genuine critique from tools optimized for fluent agreement, then, becomes the central practical problem of using them seriously, and much of what follows is an account of how I worked on that problem.

The paper's central contribution is the concept of \emph{adversarial co-thinking}: a situated, high-skill practice of using multiple GenAI tools against one another while retaining authorial judgment. Three supporting observations develop this: critique improves when tools are calibrated against real peer reviews; multi-tool disagreement exposes \emph{articulation capture}; and disclosure frameworks may need to evolve beyond tool-listing.

\vspace*{-4pt}
\section{Iterating Toward Adversarial Co-Thinking}\label{sec:practice}

\textit{Adversarial co-thinking emerged from an iterative process of learning how the tools actually behave---and what it takes to get critique out of them rather than reassurance.} I currently live ten hours behind my collaborators, which makes real-time exchange logistically constrained. I frequently had ideas that needed refinement but no colleague was available in real time. GenAI tools partially filled this gap: I treated them as enthusiastic interns who are neither particularly knowledgeable nor critical, approaching every suggestion with skepticism, and using the exchange as a way of thinking out loud. The workshop call raises concerns about disruption of ``flow states'' in deep work~\cite{sandhaus2026interrogating}; my experience points in the opposite direction. Having a sparring partner available kept ideas in motion past impasses---the bar for usefulness was friction, not insight. Even bad ideas can spark better ones, which is what Suh et al.~\cite{suh2025devils} mean by ``devil's advocates'' rather than passive translators.

However, AI tools are, by default, sycophantic. Sun et al.~\cite{sun2026friendly} found that when LLMs combine complimentary demeanor with opinion alignment, perceived authenticity and trust actually decrease. Across all three platforms, I received praise for my argument's clarity without identification of genuine weaknesses. Multiple times, the tools also confused draft versions and consistently praised whichever they took to be the latest. In my experience, the most rigorous critical feedback has consistently come from peer reviews at reputable HCI venues, substantially more so than from any AI tool. This led to a practical intervention.

\textit{The first intervention was calibration: giving each tool an external standard of what serious critique looks like.} Across my recent work, I inserted peer reviews I had received from reputable HCI venues into each AI system and asked it to review my drafts through the same critical lens, as ACM policy allows authors to use and share the text of reviews they receive~\cite{acmpeerreview2025}---although routing such text through commercial models, with varying retention policies, raises its own questions. The quality of feedback improved markedly once the tools had a concrete model of what critical engagement looked like. Alongside calibration, I used persona prompting: instructing each tool to take on the perspective of an experienced reviewer before evaluating drafts. The two techniques together moved feedback away from praise and broad clarity remarks toward identification of recurring weaknesses such as overclaiming, insufficient positioning, and unclear contribution boundaries. Without that calibration, AI feedback converged on approval rather than challenge. This is consistent with Chen et al.'s~\cite{chen2025peer} empirical finding that ChatGPT assistance reduced reviewer workload but did not improve review quality, suggesting that uncalibrated AI assistance does not by itself produce more substantive feedback. Technically, this calibration is few-shot prompting; it can also be read as an LLM-as-judge approach using peer reviews rather than generic rubrics.

When I fed the reviews into the AI tools and asked them to evaluate new drafts with the same critical stance, the AI was able to recognize recurring patterns across all my writing---including drafts the reviewers had never seen. The AI did not diagnose the problems---the human reviewers did. But once calibrated with their standards, AI helped me see where the patterns recurred, and I was able to address them. This raises an open question about scalability: the technique as I used it drew on reviews of my own prior work. Whether calibration with a shared corpus of anonymized substantive reviews---rather than a researcher's own---would be similarly effective is an open empirical question. The technique also has a saturation point: once a manuscript had essentially addressed all identified issues, AI suggestions began to degrade rather than improve the text. The next step had to come from elsewhere.

\textit{The next step was adversarial: setting the tools against one another so that their disagreements did the critical work calibration alone no longer could.} Adversarial co-thinking, in the form practiced here, was triangulation rather than delegation. I posed the same question to two or three tools, compared responses myself, and at times fed one tool's output to another for critique---but the tools never communicated directly. All routing went through me, which is also where each tool's reaction to a competitor's framing became visible: Claude, notably, often responded to opposing feedback with what read as mild offense, defending its earlier formulations rather than engaging the substance.

\vspace*{-4pt}
\section{Adversarial Co-Thinking in Practice}\label{sec:writing}

Each tool received the same starting point. I provided the same materials---the workshop call, a website from an adjacent past workshop, and a \LaTeX{} template---and the same prompt: \emph{``Can you draft me a position paper based on the work we have done together so far?''} AI outputs vary across sessions and model versions\footnote{Models used during the writing of this paper: Gemini 3 Flash with Thinking enabled, GPT-5 in Thinking mode, and Claude Opus 4.7 Adaptive.}, so the patterns described below should be expected to hold only approximately for other versions of the same tools. Even so, sycophancy, framing drift, and tool-specific defaults recurred across many context windows over the weeks this paper was iterated.

The three drafts differed substantially. Gemini introduced dramatic concepts like ``Interpretive Sovereignty''---overclaiming well beyond what my experience warranted. ChatGPT proposed ``articulation infrastructures'' as a theoretical lens, conceptually tidy but almost entirely impersonal, reading as an abstract theory paper rather than a reflexive account. Claude stayed closest to my workflow, centering evaluative judgment and citing relevant literature, but omitted the points I considered most essential. None captured what I wanted to say; all defaulted to theoretical abstraction.

A concrete moment illustrates how the evaluative work actually proceeded. After analyzing the initial drafts, I provided each tool with my own notes specifying the intended argument and asked it to revise. ChatGPT's first response opened with: ``Now this is a much clearer direction---and honestly, your material is stronger than any of the individual drafts.'' The substance that followed, however, subordinated my points to its ``articulation infrastructure'' framework---my experience became an illustration for its theory. Gemini, working from the same notes, inflated every point and coined new terminology, borrowing dramatic language from its first draft without attribution. Claude restructured around my priorities more willingly but still required substantial rewriting. The evaluative question at this point was not which draft was best, but which moves served the argument I had specified and which were the tool's own framing reasserting itself.

Two things from the AI drafts proved worth keeping. ChatGPT coined \emph{articulation capture} for a pattern across all three tools: the risk that a writer working with GenAI will absorb a tool's preferred framings without recognizing them as such---mistaking the tool's house style for their own emerging position. This extends Jakesch et al.'s~\cite{jakesch2023cowriting} finding that co-writing with opinionated language models shifts users' expressed and held views without awareness, from propositional persuasion to absorption of stylistic and structural defaults. Claude's draft, separately, introduced \emph{adversarial co-thinking}---now this paper's central frame and title. Neither term was an idea the AI had developed; both were concise labels for patterns I was already observing, and the criterion for keeping them was whether they served a vision of the paper I had already formed. Recognizing them as signal was itself the evaluative work this paper describes---particularly valuable to a non-native English speaker who can explain a concept in ten words but often cannot reach the term that captures it cleanly. Claude also showed the inverse risk: unprompted, it offered two narrowing directions---adversarial co-thinking, or an equity framing around access and the supervision gap. I chose the first; Claude's reorganization drifted toward the second---access became the opening, articulation capture a downstream symptom. The drift was visible only on reading through; reverting was easy because no real writing had been done. Fluent reorganization can naturalize a framing the writer has explicitly rejected, and adversarial co-thinking, by canceling divergent house styles, is the most direct mitigation against \emph{framing capture}.

For this paper, I deliberately set aside the boundaries that I had imposed on AI use in my earlier work. I wanted to apply adversarial co-thinking fully, at every step from initial drafting onward, without holding back on the basis of habit or caution---to see how far the workflow could go. The limits of what current GenAI tools can and cannot achieve become visible only through cases that push further than convention allows. Without such testing, community norms will remain shaped by caution rather than evidence. I believe that my own voice remains legible---the points made here are the points I intended to make. However, the literary quality---paragraph structure, sentence rhythm, section organization---was shaped by the tool in ways difficult to fully disentangle from my own contribution. This is \emph{stylistic capture} at sentence level: tools may converge on a shared register even as they diverge on framing, and this paper's prose is no exception. That entanglement is precisely why I have come to treat one practice as non-negotiable: the author must always review any text that has passed through AI in its final form, sentence by sentence, word by word, printed on paper. Even when scoped narrowly, models may quietly rewrite emphasis, soften a claim, or substitute terms~\cite{sarkar2024llms}, and the user cannot see how those decisions get made.

During this paper's development, I showed full versions to people I trust to push back rather than reassure. The same drafts that AI tools had praised consistently drew substantive critique from these readers, pushing the paper further than any AI exchange had. Without that external check, AI's praise can produce a false sense of accomplishment. Articulation capture extended into stance itself when Claude inserted, unprompted, a caveat hedging my judgment that human feedback improved the paper. It can operate on confidence and epistemic position, not only on style and concepts.

\section{What Adversarial Co-Thinking Reveals}\label{sec:reveals}

The workshop call raises concerns about de-skilling~\cite{sandhaus2026interrogating}. Multi-tool workflows do not eliminate skill requirements; they shift them. Evaluative judgment becomes critical competence: distinguishing AI outputs that are sound from those that merely sound right---recognizing overclaiming, unattributed borrowing of conceptual language, flattened arguments, and praise that masquerades as critique. A practical corollary was refusing to retain any sentence I could not paraphrase back---a counter to the illusion of explanatory depth He et al.~\cite{he2025conversational} identify. Adversarial co-thinking works only when the author has the domain-specific knowledge to weigh competing outputs; Lee et al.~\cite{lee2025impact} found that critical thinking diminishes as confidence in GenAI increases, while self-confidence in one's own domain preserves critical thinking. AI-assisted reframing is not co-authorship: the argument was redirected more than once in directions none of the tools proposed, and every such decision came from me. What the AI offered was something to react against, partially filling a function an in-person research community might otherwise have filled; the reaction itself was also formative, surfacing recurring habits in my own writing that I can now identify unaided. The co-thinking was genuine but asymmetric: the tools were not thinking partners, but their outputs added threads to a train of thought that remained mine to develop, reject, or redirect. This echoes Manuvinakurike et al.'s~\cite{manuvinakurike2025thoughts} ``explanations without explainability''---text that appears reasoned but does not reliably guide toward correct outputs. Without domain expertise, fluent reasoning reads as competent reasoning, with no internal reference to test it against. Adversarial co-thinking is a high-skill epistemic practice: it amplifies expertise but masks its absence.

One assumption worth challenging is that GenAI tools make academic writing faster. For this paper, they did not: the time was roughly comparable to writing without AI. But efficiency was never the relevant measure. The iterative exchange with the tools is the subject of this paper---without it, there would be no account to write, and the thinking it documents would not have taken the shape it did. The exchange also produced literary quality I could not have managed unaided as a non-native English speaker. The case for adversarial co-thinking is therefore not efficiency but evaluative leverage: researchers without ready access to substantive critique elsewhere may find the trade worthwhile; those embedded in active critical communities may not, and the workflow should not be defended as if speed were among its benefits.

Current disclosure practices typically list tools used and state that authors assume full responsibility~\cite{acm2023policy}. This obscures the texture of the collaboration---which suggestions the author took, rejected, or revised. Since LLMs' confident tone is not evidence of reasoning quality~\cite{sarkar2024llms}, tool-listing alone is insufficient.

\section{Implications for the Workshop}\label{sec:implications}

Four propositions are offered for workshop discussion, framed as questions the single situated case can motivate but not settle.

\textbf{Proposition 1: Do multi-tool workflows better support professional autonomy than single-tool reliance?} Single-tool use exposes the writer to only one set of framings, narrowing over time the range of expression they reach for. Multi-tool use introduces friction and forces comparative judgment, a self-imposed cognitive forcing function~\cite{bucinca2021trust} that may reduce overreliance on any single fluent output. This does not mean more AI is always better; whether multi-tool use preserves more autonomy than careful single-tool use, for researchers with the expertise to evaluate competing outputs, remains an empirical question.

\textbf{Proposition 2: Calibration with real review standards extends AI's usefulness to both authors and reviewers.} For isolated researchers, the more pressing risk is not losing existing skills but never acquiring the contextual knowledge---venue conventions, what travels and what does not---that well-connected researchers absorb from their environment. This is distinct from substantive domain expertise (Section~\ref{sec:reveals}), which remains the researcher's irreducible contribution. Contextual knowledge is more tacit and more amenable to partial substitution: feeding real peer reviews into AI tools transformed feedback from generic approval to actionable critique, suggesting a design direction in conference-specific review modes. For authors, this could help researchers unfamiliar with venue conventions receive useful feedback before submission. For reviewers, calibrating AI assistance with substantive prior reviews of a venue could move past Chen et al.'s~\cite{chen2025peer} finding that ChatGPT reduced workload without improving review quality.

\textbf{Proposition 3: For whom is the multi-tool workflow actually available?} Adversarial co-thinking presupposes access to multiple GenAI platforms with enough usage capacity to sustain comparison. In my case, this meant paid subscriptions across three providers plus a Max-tier upgrade---costs beyond the reach of many researchers. Native English speakers already benefit when disseminating research; premium AI access compounds that asymmetry. Cost is not the only constraint, though. Even where budget exists, institutional data governance policies can steer researchers toward institutionally-provisioned single tools, and researchers' autonomy to deviate from those defaults varies. Whether multi-tool triangulation is qualitatively different from careful single-tool use is an empirical question worth the workshop's attention. If it is, the question of who can actually run the workflow---and what determines that---is one the field should take up.

\textbf{Proposition 4: What level of AI disclosure is workable in practice?} Listing tool names and asserting responsibility does not capture the texture of AI-augmented work, but comprehensive process-level disclosure---accounting for every suggestion taken, rejected, or revised---could easily become longer than the paper itself, and applies a standard not consistently expected of human collaborators. The narrower claim is that disclosure should at least describe where AI shaped framing or argument (the domain where articulation capture operates), distinct from where it served only as language polish. Where exactly that line is drawn---and whether the AI/human-collaborator distinction holds---is a question this paper raises but cannot resolve alone.

\section{Conclusion}

This position paper describes adversarial co-thinking with multiple GenAI tools as it emerged during recent academic writing in HCI. The tools are useful sparring partners, structural critics, and sources of alternative framings, but they are not intellectual peers. The tools did not determine this paper's intellectual direction; their useful contributions were labels, alternative framings, and objections to react against. They do not provide critical feedback at the level of scientific rigor unless explicitly calibrated. And they are not equally available to all researchers, given the cost of maintaining multiple subscriptions. The challenge is no longer whether to use AI---AI will be used. The challenge is how to use it without letting it silently define the boundaries of what we are able to think and say, and how to communicate that use honestly in the work we publish.

\begin{acks}
\textbf{AI Use Statement:} This paper was developed through the adversarial co-thinking process it describes. Initial drafts from Claude, ChatGPT, and Gemini failed to capture the author's intended argument, though three formulations were retained where they captured patterns the author was already observing: \emph{adversarial co-thinking} (Claude), \emph{articulation capture} (ChatGPT), and the phrase ``amplifies expertise but masks its absence'' (Gemini). The author produced notes specifying the intended argument, made all decisions about framing and final wording, and wrote the paper iteratively with Claude, aided by the other tools. During revision, hostile-reviewer simulations were requested from all three tools against the current draft, and their convergent critiques guided the revisions.

\textbf{Critical Readers:} Karolina Drobotowicz, Taneli Mielik\"ainen, Kari Pulli, Antti Rannisto, and Petri Vuorimaa.
\end{acks}

\bibliographystyle{ACM-Reference-Format}
\bibliography{references}

@inproceedings{sandhaus2026interrogating,
  author    = {Sandhaus, Hauke and Imteyaz, Kashif and Almutairi, Mohammed
               and Prajod, Pooja and Ramesh, Divya and Savage, Saiph
               and Yang, Qian and Muller, Michael},
  title     = {Interrogating {GenAI} Augmentation for {CHIworkers}:
               Strategies for Professional Autonomy and Accountability},
  year      = {2026},
  isbn      = {979-8-4007-2598-2/2026/06},
  publisher = {Association for Computing Machinery},
  address   = {New York, NY, USA},
  url       = {https://doi.org/10.1145/3805029.3818271},
  doi       = {10.1145/3805029.3818271},
  booktitle = {Adjunct Proceedings of the 5th Annual Symposium on
               Human-Computer Interaction for Work},
  series    = {CHIWORK '26},
  location  = {Linz, Austria},
  month     = jun,
}

@article{desjardins2021firstperson,
  author = {Desjardins, Audrey and Tomico, Oscar and Lucero, Andr{\'e}s and Cecchinato, Marta E. and Neustaedter, Carman},
  title = {Introduction to the Special Issue on First-Person Methods in {HCI}},
  journal = {ACM Transactions on Computer-Human Interaction},
  year = {2021},
  volume = {28},
  number = {6},
  articleno = {37},
  numpages = {12},
  doi = {10.1145/3492342}
}

@inproceedings{chen2025peer,
  author = {Chen, Shiping and Brumby, Duncan and Cox, Anna},
  title = {Envisioning the Future of Peer Review: Investigating 
           {LLM}-Assisted Reviewing Using {ChatGPT} as a Case Study},
  booktitle = {Proceedings of the 4th Annual Symposium on 
               Human-Computer Interaction for Work (CHIWORK '25)},
  year = {2025},
  publisher = {ACM},
  address = {Amsterdam, Netherlands},
  doi = {10.1145/3729176.3729196}
}

@inproceedings{sarkar2024llms,
  author    = {Sarkar, Advait},
  title     = {Large Language Models Cannot Explain Themselves},
  booktitle = {Proceedings of the 4th ACM CHI Workshop on
               Human-Centered Explainable AI (HCXAI '24)},
  year      = {2024},
  note      = {arXiv:2405.04382},
}

@inproceedings{suh2025devils,
  author    = {Suh, Ashley and Alperin, Kenneth and Li, Harry and
               Gomez, Steven R},
  title     = {Don't Just Translate, Agitate: Using Large Language Models
               as Devil's Advocates for {AI} Explanations},
  booktitle = {Proceedings of the 5th ACM CHI Workshop on
               Human-Centered Explainable AI (HCXAI '25)},
  year      = {2025},
  doi       = {10.5281/zenodo.15170455},
}

@inproceedings{manuvinakurike2025thoughts,
  author    = {Manuvinakurike, Ramesh and Moss, Emanuel and
               Watkins, Elizabeth Anne and Sahay, Saurav and
               Raffa, Giuseppe and Nachman, Lama},
  title     = {Thoughts without Thinking: Reconsidering the Explanatory
               Value of Chain-of-Thought Reasoning in {LLMs} through
               Agentic Pipelines},
  booktitle = {Proceedings of the 5th ACM CHI Workshop on
               Human-Centered Explainable AI (HCXAI '25)},
  year      = {2025},
  doi       = {10.5281/zenodo.15170393},
}

@inproceedings{he2025conversational,
  author = {He, Gaole and Aishwarya, Nilay and Gadiraju, Ujwal},
  title = {Is Conversational {XAI} All You Need? {H}uman-{AI} Decision Making With a Conversational {XAI} Assistant},
  booktitle = {Proceedings of the 30th International Conference on Intelligent User Interfaces (IUI '25)},
  year = {2025},
  publisher = {ACM},
  address = {Cagliari, Italy},
  doi = {10.1145/3708359.3712133}
}

@inproceedings{bucinca2021trust,
  author = {Bu{\c{c}}in{\c{c}}a, Zana and Malaya, Maja Barbara and Gajos, Krzysztof Z.},
  title = {To Trust or to Think: Cognitive Forcing Functions Can Reduce Overreliance on {AI} in {AI}-assisted Decision-making},
  journal = {Proceedings of the ACM on Human-Computer Interaction},
  year = {2021},
  volume = {5},
  number = {CSCW1},
  articleno = {188},
  numpages = {21},
  doi = {10.1145/3449287}
}

@inproceedings{jakesch2023cowriting,
  author    = {Jakesch, Maurice and Bhat, Advait and Buschek, Daniel and Zalmanson, Lior and Naaman, Mor},
  title     = {Co-Writing with Opinionated Language Models Affects Users' Views},
  year      = {2023},
  isbn      = {9781450394215},
  publisher = {Association for Computing Machinery},
  address   = {New York, NY, USA},
  url       = {https://doi.org/10.1145/3544548.3581196},
  doi       = {10.1145/3544548.3581196},
  booktitle = {Proceedings of the 2023 CHI Conference on Human Factors in Computing Systems},
  articleno = {111},
  numpages  = {15},
  location  = {Hamburg, Germany},
  series    = {CHI '23}
}

@inproceedings{lee2025impact,
  author = {Lee, Hao-Ping (Hank) and Sarkar, Advait and Tankelevitch, Lev and Drosos, Ian and Rintel, Sean and Banks, Richard and Wilson, Nicholas},
  title = {The Impact of Generative {AI} on Critical Thinking: Self-Reported Reductions in Cognitive Effort and Confidence Effects From a Survey of Knowledge Workers},
  booktitle = {Proceedings of the 2025 CHI Conference on Human Factors in Computing Systems},
  year = {2025},
  publisher = {ACM},
  address = {Yokohama, Japan},
  pages = {1--22},
  doi = {10.1145/3706598.3713778}
}

@inproceedings{sun2026friendly,
  author = {Sun, Yuan and others},
  title = {Be Friendly, Not Friends: How {LLM} Sycophancy Shapes User Trust},
  booktitle = {Proceedings of the 2026 CHI Conference on Human Factors in Computing Systems},
  year = {2026},
  publisher = {ACM},
  address = {Barcelona, Spain},
  doi = {10.1145/3772318.3791079}
}

@misc{acmpeerreview2025,
  author = {{ACM}},
  title = {Peer Review Policy: Frequently Asked Questions},
  year = {2025},
  url = {https://www.acm.org/publications/policies/peer-review-faq},
  note = {Accessed: 2026-04-22}
}

@misc{acm2023policy,
  author    = {{ACM}},
  title     = {Policy on Authorship},
  year      = {2023},
  howpublished = {\url{https://www.acm.org/publications/policies/new-acm-policy-on-authorship}},
  note      = {Accessed: 2026-04-20},
}

\end{document}